%% file: qfabric-paper.tex
\documentclass[conference]{IEEEtran}

\usepackage[T1]{fontenc}
\usepackage[utf8]{inputenc}
\usepackage{amsmath,amssymb}
\usepackage{booktabs}
\usepackage{graphicx}
\usepackage{microtype}
\usepackage{multirow}
\usepackage{url}
\usepackage{xspace}
\usepackage{xcolor}
\usepackage{tikz}
\usetikzlibrary{arrows.meta,positioning,shapes.geometric,fit}
\usepackage{pgfplots}
\usepackage{pgfplotstable}
\pgfplotsset{compat=1.18}
\usepackage{dblfloatfix}
\usepackage{placeins}
\usepackage[hidelinks]{hyperref}

\input{generated/metrics.tex}
\input{generated/interconnect-metrics.tex}
\input{generated/stage12-metrics.tex}

\definecolor{qblue}{RGB}{34,92,150}
\definecolor{qorange}{RGB}{222,119,38}
\definecolor{qgreen}{RGB}{52,128,73}
\definecolor{qred}{RGB}{178,55,55}
\newcommand{\system}{QFabric\xspace}
\hypersetup{
  pdftitle={QFabric: Guarded Cross-Domain Task Recovery and Its Limits on a Linux--RTOS Board},
  pdfauthor={Ayushman Bhattacharya and Anwesha Chakraborty},
  pdfkeywords={heterogeneous embedded systems, asymmetric multiprocessing, Linux, RTOS, interconnect characterization, timing contracts}
}

\title{QFabric: Guarded Cross-Domain Task Recovery\\and Its Limits on a Linux--RTOS Board}

\author{\IEEEauthorblockN{Ayushman Bhattacharya}
\IEEEauthorblockA{\textit{Pollinations AI}\\
\href{mailto:ayushman.elixpo@gmail.com}{ayushman.elixpo@gmail.com}}
\and
\IEEEauthorblockN{Anwesha Chakraborty}
\IEEEauthorblockA{\textit{Dept. CSE, JIS University}\\
\href{mailto:anwesha.elixpo@gmail.com}{anwesha.elixpo@gmail.com}}}

\begin{document}
\maketitle

\begin{abstract}
Heterogeneous embedded boards pair a Linux application processor with a
real-time microcontroller (MCU), which suggests moving a task between the two
domains when its timing degrades. We present \system, a prototype that treats
cross-domain placement as guarded recovery from a violated empirical soft
real-time contract: a bounded task is compiled for both domains, monitored in
windows with hysteresis, switched only after semantic and capacity admission
at a quiescent epoch boundary, and committed or rolled back after probation.
On an Arduino UNO Q (Cortex-A53 Linux and a Cortex-M33 running Zephyr), we
characterize the stock Bridge RPC interconnect and measure five workloads in
both domains under idle, CPU-load, and higher-priority \texttt{SCHED\_FIFO}
contention, with ordinary and real-time callers. Each MCU invocation costs
about 19~ms of dispatch regardless of compute, the MCU runs the heaviest
kernel \StageTwelveFirComputeRatio$\times$ slower, and because every call
traverses Linux userspace, Linux contention inflates the MCU path far more
than the Linux path (p99 up to \StageTwelveContentionRtPNinetyNine~ms versus
\StageTwelveContentionLinuxPNinetyNine~ms). Across \StageTwelveCells{} cells
of a four-node chain, Linux meets every contract within
\StageTwelveWorstLinuxMiss{} misses and \system correctly never migrates; a
controller without hysteresis migrates in \StageTwelveAblatedSwitches{} cells,
raising p95 by up to \StageTwelveAblationWorstRatio$\times$, and probation
rolls back \StageTwelveAblatedRollbacks{} of those moves. On Linux-dispatched
heterogeneous boards, migration therefore cannot isolate a task from Linux
interference, and recovery guards matter chiefly by preventing harmful moves.
This paper supersedes an earlier artifact release whose reported improvement
arose from measurement defects that we identify. All source, raw traces, and
generators are public.
\end{abstract}


\section{Introduction}
\label{sec:intro}
Embedded platforms increasingly integrate two execution environments on one
board: a Linux application processor for libraries, networking, and developer
productivity, and a real-time microcontroller for isolation and predictable
execution~\cite{arduinoUnoQ,openampProject}. A task that can run in either
domain faces a placement choice whose consequences extend beyond compute
time. Crossing the boundary adds serialization and transport, contention
reshapes tail latency, and a move may be illegal for stateful or
Linux-specific operations.

A natural idea is to treat the microcontroller as a refuge: when a task's
timing degrades on Linux, move it to the real-time domain. Classical
heterogeneous scheduling maps task graphs statically from cost
estimates~\cite{topcuoglu2002heft}; feedback scheduling adapts resources to
measured error~\cite{lu2002feedback}; and runtime verification makes timing
violations observable~\cite{reinbacher2014runtime,nandi2015contracts}. What
remains unclear on asymmetric Linux--RTOS boards is whether such a move can
actually help, and how a runtime should decide when a faster prediction is not
sufficient permission to migrate.

We study this question with \system, a deliberately narrow prototype on the
Arduino UNO Q. A portable task declaration is compiled for both domains,
profiled in each, and governed by an empirical contract. A recovery controller
applies hysteresis to complete observation windows, admits a candidate only if
semantic, capacity, and predicted-cost conditions hold, drains in-flight work
at an epoch boundary, and commits the move only if post-switch probation
confirms the contract; otherwise it rolls back.

The central finding is a boundary, not a speedup. On this board every MCU
invocation pays a fixed dispatch cost of about 19~ms, and because that
invocation is issued and relayed by Linux processes, Linux interference
degrades the MCU path more than the Linux path. Migration therefore never
improves a contract in our measurements; \system's guards earn their place by
refusing the moves that a naive controller would make. An earlier artifact
release~\cite{qfabric2026v101} reported a 36.6\% improvement from recovery;
Section~\ref{sec:correction} identifies the measurement defects behind that
result and shows that it does not hold.

\textbf{Contributions.} This paper makes four contributions. First, it
presents a guarded recovery protocol that combines a bounded wire ABI,
effect-based semantic admission, quiescent epoch switching, and
probation-based commit or rollback (Section~\ref{sec:design}). Second, it
characterizes the UNO Q Linux--MCU interconnect across payload size,
concurrency, load, and frequency governor (Section~\ref{sec:interconnect}).
Third, it reports a matched-load measurement study of five workloads in both
domains under three load conditions and two caller policies, decomposing each
call into compute and residual dispatch cost (Section~\ref{sec:percall}).
Fourth, it evaluates the controller's decisions on a four-node chain and shows
through an ablation that hysteresis and probation prevent harmful migrations
(Sections~\ref{sec:chain}--\ref{sec:guards}).

\textbf{Positioning.} Approaches to cross-domain placement differ in when the
assignment of software to execution resources is fixed, what triggers a
reassignment, and whether a reassignment must prove its benefit before it is
kept. Static mapping binds at design time~\cite{topcuoglu2002heft};
asymmetric-multiprocessing frameworks and partitioning hypervisors bind at
configuration time~\cite{openampProject,ramsauer2017jailhouse,martins2020bao};
and feedback scheduling, mode-change protocols, and migration runtimes rebind
at runtime but keep a move once it is made
\cite{lu2002feedback,real2004mode,goraczko2006comos,rouxel2021yasmin}.
\system rebinds at runtime only after an observed contract violation, and
keeps a move only after observing that it helped.
Table~\ref{tab:related} in Section~\ref{sec:related} compares these
approaches in detail.

The rest of the paper is organized as follows.
Section~\ref{sec:background} describes the platform and system model,
Section~\ref{sec:related} reviews and compares related work,
Section~\ref{sec:design} presents the design, Section~\ref{sec:impl} the
implementation, Section~\ref{sec:method} the methodology,
Section~\ref{sec:results} the results, and Section~\ref{sec:discussion}
discusses implications and threats to validity before
Section~\ref{sec:conclusion} concludes.

\section{Background and System Model}
\label{sec:background}
\subsection{Target platform}
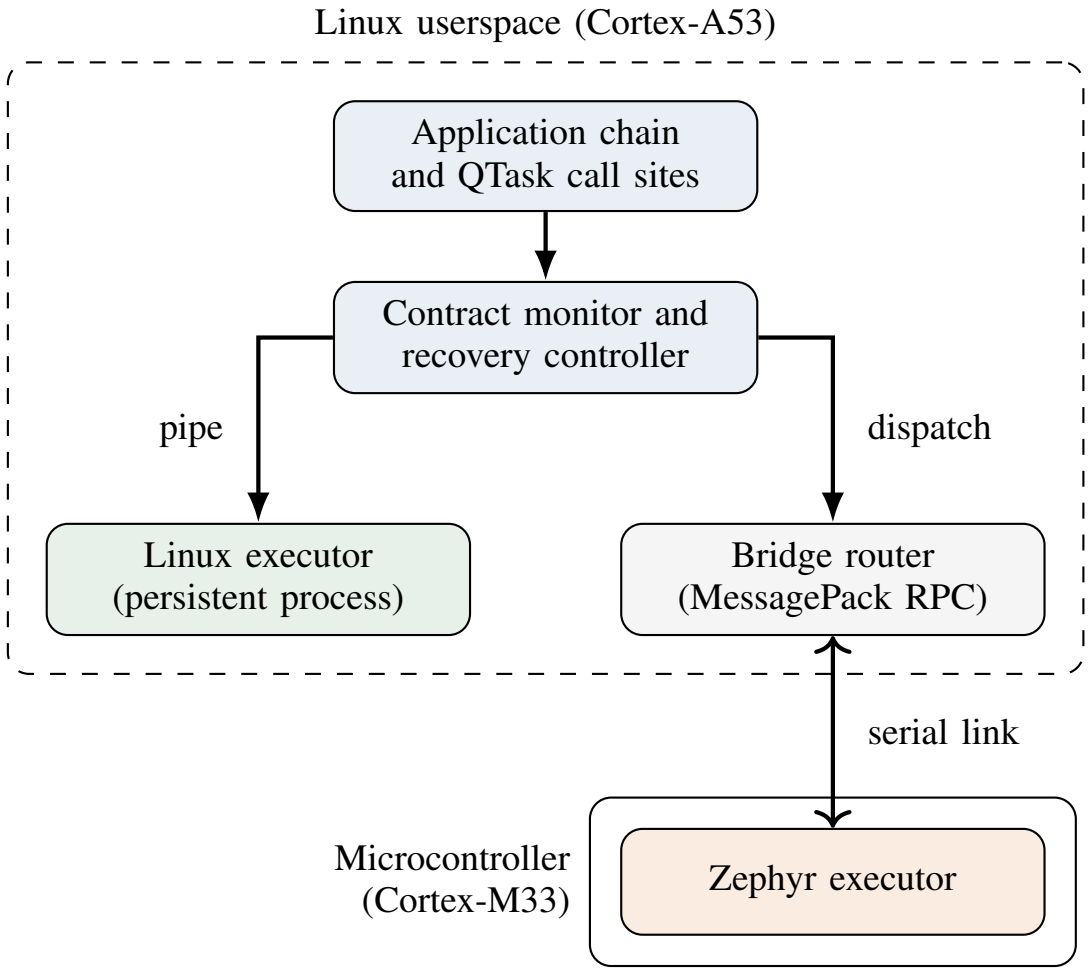
\begin{figure}[t]
\centering
\begin{tikzpicture}[
  font=\scriptsize,
  box/.style={draw,rounded corners,minimum width=28mm,minimum height=7mm,align=center},
  flow/.style={-{Latex[length=2mm]},thick}
]
\node[box,fill=qblue!10] (app) at (0,0) {Application chain\\and QTask call sites};
\node[box,fill=qblue!10] (ctl) at (0,-12mm) {Contract monitor and\\recovery controller};
\node[box,fill=qgreen!12] (linux) at (-19mm,-28mm) {Linux executor\\(persistent process)};
\node[box,fill=gray!8] (router) at (19mm,-28mm) {Bridge router\\(MessagePack RPC)};
\node[draw,dashed,rounded corners,fit=(app)(ctl)(linux)(router),inner sep=2.5mm,
      label={[font=\scriptsize]above:{Linux userspace (Cortex-A53)}}] {};
\node[box,fill=qorange!14] (rt) at (19mm,-48mm) {Zephyr executor};
\node[draw,rounded corners,fit=(rt),inner sep=2mm,
      label={[font=\scriptsize,align=right]left:{Microcontroller\\(Cortex-M33)}}] {};
\draw[flow] (app) -- (ctl);
\draw[flow] (ctl.west) -| node[left=1mm,pos=0.75]{pipe} (linux.north);
\draw[flow] (ctl.east) -| node[right=1mm,pos=0.75]{dispatch} (router.north);
\draw[flow,<->] (router) -- node[right=1mm]{serial link} (rt);
\end{tikzpicture}
\caption{\system on the UNO Q. The controller, dispatcher, and Bridge router
run in Linux userspace; each task has a native implementation in both
domains.}
\label{fig:architecture}
\end{figure}

The Arduino UNO Q combines a quad-core Arm Cortex-A53 application processor
running Debian Linux with an STM32U585 Cortex-M33 microcontroller running
Zephyr~\cite{arduinoUnoQ}. Arduino's Bridge stack exposes MessagePack remote
procedure calls in both directions through a Linux-resident router
process~\cite{arduinoRouterBridge}. The arrangement resembles
remote-processor systems built on \texttt{remoteproc} and \texttt{rpmsg}
\cite{linuxRemoteproc,linuxRpmsg}, but we use the stock UNO Q stack rather
than modifying the transport. Fig.~\ref{fig:architecture} shows how \system
sits on this platform: the policy, the dispatcher, and the Bridge router all
execute in Linux userspace, so every MCU invocation begins and ends there.

\subsection{System model}
An application is a chain $G=(V,E)$ of tasks. Each task $v\in V$ may have a
Linux implementation, an MCU implementation, or both, together with
fixed-width request and response types, a numeric identifier, and an effect
class. Only pure, bounded tasks are migration eligible. A placement assigns a
domain $d(v)\in\{\mathrm{L},\mathrm{M}\}$ to every task.

Every invocation is issued from Linux userspace, so the latency observed for
task $v$ in domain $d$ decomposes into local compute and a residual that
captures dispatch, transport, and scheduling delay:
\begin{equation}
  L_{v,d} = C_{v,d} + R_{v,d}.
  \label{eq:decomposition}
\end{equation}
For the MCU, $R$ includes MessagePack encoding, the Linux router, the serial
link, and the return path; for Linux it is the cost of a pipe round trip to a
persistent executor. A chain observation costs
$L_G=\sum_{v\in V} L_{v,d(v)}$, and the controller predicts it from profiled
95th percentiles:
\begin{equation}
  \widehat{C}_G(d) = \sum_{v\in V} q_{.95}(v,d(v)).
  \label{eq:cost}
\end{equation}
This is a transparent heuristic, not a worst-case execution-time bound.
Because each MCU sample already includes its round trip from Linux, no
separate boundary term is added.

For a window $w$ of $n$ invocations with outcomes $o_i$ and latencies $L_i$,
the observed miss rate against a deadline $D$ is
\begin{equation}
  \widehat{m}_w=\frac{1}{n}\sum_{i=1}^{n}
  \mathbf{1}\!\left[o_i\ne\mathrm{ok}\ \lor\ L_i>D\right],
  \label{eq:miss}
\end{equation}
and the window violates the contract when $\widehat{m}_w>\epsilon$. The
contract is an empirical soft real-time objective.

\subsection{Research questions}
We ask: \textbf{RQ1}, what does the Linux--MCU interconnect cost, and how does
that cost respond to payload, concurrency, load, and frequency policy?
\textbf{RQ2}, under matched load conditions, how do the two domains compare
per call, and where does the time go? \textbf{RQ3}, on a four-node chain, do
contract violations arise, and does the controller decide correctly against
deployable baselines and an oracle? \textbf{RQ4}, are hysteresis, admission,
and probation necessary? \textbf{RQ5}, what does the mechanism itself cost?

\section{Related Work}
\label{sec:related}
We review five lines of work that bear on cross-domain recovery and compare
representative systems in Table~\ref{tab:related}. Its first three columns
follow the dimensions introduced in Section~\ref{sec:intro}: binding time,
trigger, and the unit of isolation; the remaining columns record whether a
property is central to the cited work, not whether it could be added.

\begin{table*}[t]
\caption{Comparison of \system with related work. ``Semantic guard'' means
effect- or state-based migration eligibility; ``verified switch'' means an
observed post-switch check decides whether the move is kept.}
\label{tab:related}
\centering
\scriptsize
\setlength{\tabcolsep}{3.2pt}
\begin{tabular}{llllcccc}
\toprule
Work & Binding time & Trigger & Isolation unit & Semantic guard & RT admission & Verified switch & Linux--RTOS board \\
\midrule
HEFT~\cite{topcuoglu2002heft} & design time & cost model & processor & -- & -- & -- & -- \\
Feedback scheduling~\cite{lu2002feedback} & runtime & control error & resource share & -- & model-specific & -- & -- \\
Stochastic contracts~\cite{nandi2015contracts} & runtime & contract monitor & component & -- & -- & -- & -- \\
Mode changes~\cite{real2004mode,vestal2007preemptive} & runtime & mode or budget overrun & task set & -- & \checkmark & -- & -- \\
RT task migration~\cite{pourmohseni2020migration} & runtime & resource management & core & -- & \checkmark & -- & -- \\
CoMOS~\cite{goraczko2006comos} & runtime & runtime system & processor & -- & -- & -- & -- \\
YASMIN~\cite{rouxel2021yasmin} & runtime & scheduler & processor & -- & -- & -- & -- \\
Awkernel~\cite{yano2026awkernel} & runtime & DAG scheduler & core & -- & -- & -- & -- \\
Jailhouse / Bao~\cite{ramsauer2017jailhouse,martins2020bao} & configuration & -- & partition & -- & -- & -- & -- \\
OpenAMP / rpmsg~\cite{openampProject,linuxRpmsg} & configuration & -- & core & -- & -- & -- & \checkmark \\
\system & runtime & contract violation & task & \checkmark & \checkmark & \checkmark & \checkmark \\
\bottomrule
\end{tabular}
\end{table*}

\subsection{Static and feedback-driven placement}
HEFT ranks a task graph and maps it to heterogeneous processors using
estimated computation and communication costs~\cite{topcuoglu2002heft}. It is
the natural static reference but does not monitor deployed timing. Real-time
scheduling theory provides admission and schedulability
foundations~\cite{liu1973scheduling}, and feedback scheduling adjusts
resource allocation from runtime error~\cite{lu2002feedback}; autonomic
computing generalizes the monitor--analyze--plan--execute
loop~\cite{kephart2003autonomic}. \system uses such a loop, but its actuator
is a guarded change of execution domain.

\subsection{Timing contracts and runtime verification}
Runtime verification observes real-time behavior with bounded
intrusion~\cite{reinbacher2014runtime}. Stochastic contracts provide
empirical runtime checks for component systems~\cite{nandi2015contracts},
while formal timing contracts and zone-based models support verification and
synthesis~\cite{alkhatib2016timing,camilli2018adaptive}. These works motivate
our contract semantics; our focus is the operational protocol after a
violation.

\subsection{Asymmetric multiprocessing and partitioning}
Heterogeneous SoCs that pair application cores with real-time cores are
commonly run as asymmetric multiprocessing systems, with Linux managing a
remote RTOS through remoteproc and rpmsg; OpenAMP standardizes these
interactions~\cite{openampProject} and has been evaluated on real-time
embedded SoC platforms~\cite{alonso2021openamp}. Measurements on Zynq
UltraScale+ show that inter-processor signaling can reach sub-microsecond
latency when shared memory is used directly instead of the OpenAMP
stack~\cite{coppik2025lowlatency}. An alternative to physically separate cores
is a static partitioning hypervisor such as Jailhouse~\cite{ramsauer2017jailhouse}
or Bao~\cite{martins2020bao}; their real-time behavior and interference on Arm
platforms have been compared empirically~\cite{martins2023partitioning}, and
hypervisor-based isolation has been combined with programmable-logic memory on
heterogeneous MPSoCs~\cite{gracioli2019mixed}. In these approaches the
assignment of software to cores or partitions is fixed at configuration time.
\system takes the Linux--RTOS split and its transport as given, and changes
only which side executes a declared task after a contract violation.

\subsection{Timing analysis and mixed criticality}
\system's contracts are empirical soft real-time bounds, not WCET guarantees.
Static and measurement-based WCET analysis~\cite{wilhelm2008wcet}, and
measurement-based probabilistic timing analysis in particular
\cite{cucugrosjean2012mbpta,davis2019probabilistic}, give principled ways to
derive bounds from observed executions, under assumptions that our percentile
thresholds do not establish. The mixed-criticality literature, started by
Vestal's model of tasks with several execution-time estimates of different
assurance~\cite{vestal2007preemptive} and surveyed by Burns and
Davis~\cite{burns2017survey}, also studies runtime adaptation when a
lower-assurance estimate is exceeded. That reaction is usually a criticality
mode change, a well-studied class of protocol~\cite{real2004mode}. \system's
recovery can be read as a narrow, cross-domain instance of this pattern: a
violated empirical bound triggers a guarded, verified change of placement,
not a change of criticality mode, and we make no claim of certifiable
guarantees.

\subsection{Migration across heterogeneous processors}
Real-time task migration has been studied for dynamic resource management in
many-core systems with feasibility analysis~\cite{pourmohseni2020migration},
and Linux supplies deadline-aware scheduling
mechanisms~\cite{lelli2016deadline}. CoMOS provides processor-agnostic tasks
and runtime migration across heterogeneous processors~\cite{goraczko2006comos};
YASMIN supports functionally equivalent task implementations and user-space
real-time mapping on heterogeneous COTS platforms~\cite{rouxel2021yasmin}; and
Awkernel brings DAG scheduling into an RTOS evaluated on physical
hardware~\cite{yano2026awkernel}. \system differs in its trigger (an observed
empirical-contract violation), its guards (semantic eligibility and capacity
admission), and its insistence on post-switch verification. To the best of
our search, this combination on a physical Linux--RTOS boundary is new; we
claim no novelty for placement, profiling, feedback control, or timing
contracts individually. Our measurements also bound when the combination can
help, which prior work on migration typically does not report.

\section{QFabric Design}
\label{sec:design}
Section~\ref{sec:background} showed that every invocation, in either domain,
begins in Linux userspace and that a predicted improvement is only an
estimate. \system's design therefore treats a migration as a hypothesis: it
is attempted only when the evidence and the guards allow it, and kept only
when observation confirms it.

\subsection{QTask declaration and bounded ABI}
A QTask declaration is the single source for Linux and MCU artifacts. The
generator accepts fixed-size scalars, records, and bounded arrays; it rejects
pointers, dynamic arrays, recursive types, unsupported effects, and messages
larger than the 988-byte payload budget. Versioned big-endian headers carry a
magic value, message kind, task identifier, request identifier, payload
length, and flags. Golden vectors are checked independently in Python, C++,
and on the MCU, which turns serialization bounds and effect eligibility into
auditable facts rather than conventions.

\subsection{Contracts and admission}
Profiles exclude warm-up invocations and retain raw observations; a minimum
sample count marks an estimator valid. The per-task deadline is fixed by rule
before any replay, from the idle profiles of both domains with ordinary
callers:
\begin{equation}
  D=\left\lceil 1.25\max_{d\in\{\mathrm{L},\mathrm{M}\}}
  p99^{\mathrm{idle}}(v,d)\right\rceil_{1\,\mathrm{ms}},\qquad
  D_G=|V|\,D .
  \label{eq:deadline}
\end{equation}
Taking the slower domain makes the contract feasible for both alternatives at
idle, so a violation reflects degradation rather than an infeasible target.
A candidate MCU move must also leave utilization below a reserved headroom $h$:
\begin{equation}
  U'_{\mathrm{M}}=U_{\mathrm{M}}+r_v\,q_{.95}(v,\mathrm{M})\le 1-h,
  \label{eq:admission}
\end{equation}
where $r_v$ is the invocation rate. A switch is admissible only when
\begin{align}
  A(v,d)={}&E_{\min}\land S_{\mathrm{semantic}}\land
  S_{\mathrm{admission}} \notag\\
  &\land \bigl(\widehat{C}_G(d)\le D_G\bigr)\land S_{\mathrm{cooldown}},
  \label{eq:admissible}
\end{align}
and the controller reports a stable rejection code for every failed conjunct.

\subsection{Guarded recovery protocol}
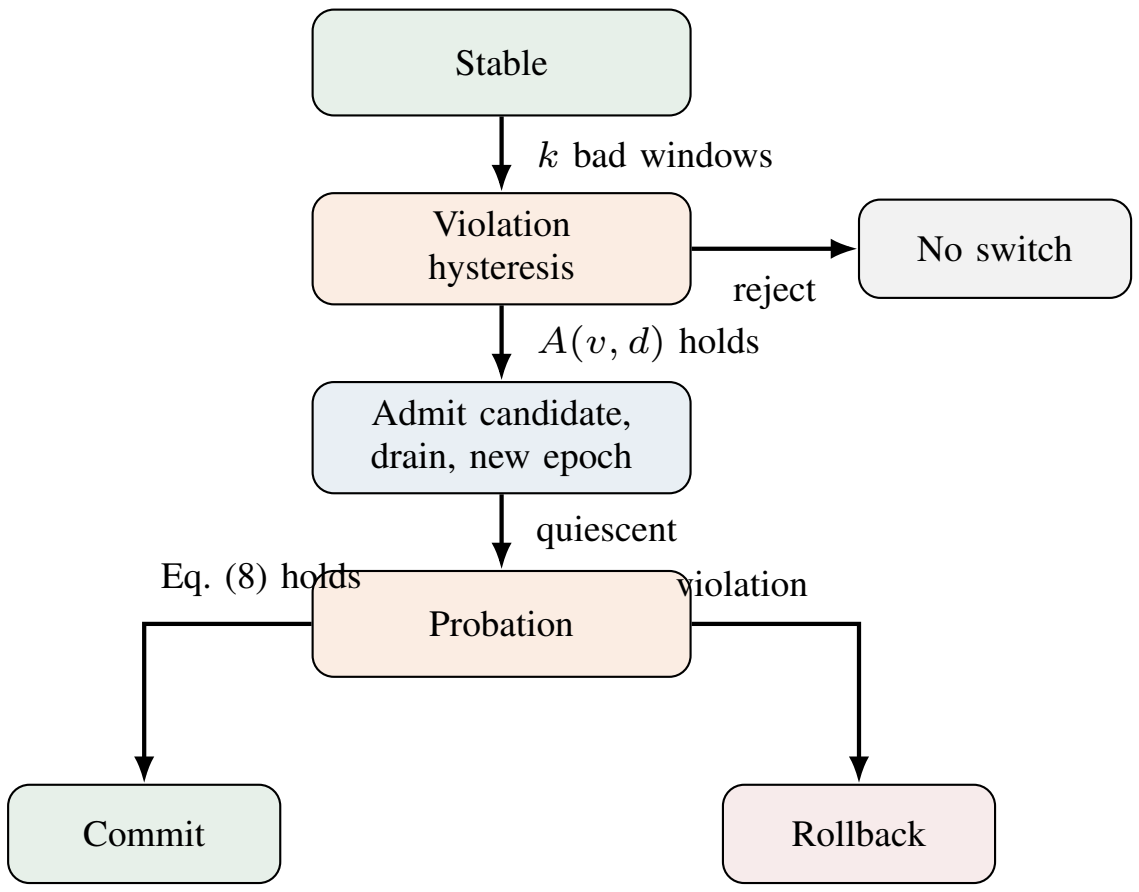
\begin{figure}[t]
\centering
\begin{tikzpicture}[
  font=\scriptsize,
  node distance=5mm and 7mm,
  state/.style={draw,rounded corners,minimum width=25mm,minimum height=7mm,align=center},
  outcome/.style={draw,rounded corners,minimum width=18mm,minimum height=6.5mm,align=center},
  arr/.style={-{Latex[length=1.8mm]},thick}
]
\node[state,fill=qgreen!12] (stable) {Stable};
\node[state,fill=qorange!13,below=of stable] (viol) {Violation\\hysteresis};
\node[state,fill=qblue!10,below=of viol] (drain) {Admit candidate,\\drain, new epoch};
\node[outcome,fill=gray!10,right=11mm of viol] (noswitch) {No switch};
\node[state,fill=qorange!13,below=of drain] (prob) {Probation};
\node[outcome,fill=qgreen!12,below left=7mm and 2mm of prob] (commit) {Commit};
\node[outcome,fill=qred!10,below right=7mm and 2mm of prob] (rollback) {Rollback};
\draw[arr] (stable) -- node[right=1mm,pos=0.5]{$k$ bad windows} (viol);
\draw[arr] (viol) -- node[right=1mm,pos=0.5]{$A(v,d)$ holds} (drain);
\draw[arr] (viol) -- node[below=0.5mm,pos=0.5]{reject} (noswitch);
\draw[arr] (drain) -- node[right=1mm,pos=0.5]{quiescent} (prob);
\draw[arr] (prob.west) -| node[above=0.5mm,pos=0.15]{Eq.~\eqref{eq:commit} holds} (commit.north);
\draw[arr] (prob.east) -| node[above=0.5mm,pos=0.15]{violation} (rollback.north);
\end{tikzpicture}
\caption{Guarded recovery. Every terminal outcome returns the monitor to
Stable.}
\label{fig:state}
\end{figure}

Fig.~\ref{fig:state} shows the recovery state machine. With windows of
$n=20$ invocations and tolerance $\epsilon=1\%$, analysis begins only after
$k$ consecutive violating windows:
\begin{equation}
  \mathrm{analyze}\iff \bigwedge_{j=0}^{k-1}\widehat{m}_{w-j}>\epsilon,
  \qquad k=3.
  \label{eq:hysteresis}
\end{equation}
If no candidate satisfies Eq.~\eqref{eq:admissible}, placement is unchanged.
Otherwise the dispatcher stops accepting work, waits until no invocation is
in flight, increments a routing epoch, and changes the active domain. The move
is committed only if the pooled miss rate over $P=6$ probation windows meets
the contract,
\begin{equation}
  \mathrm{commit}\iff \widehat{m}_{\mathrm{prob}}\le\epsilon,
  \label{eq:commit}
\end{equation}
and is rolled back to the previous epoch and domain otherwise. A prediction
thus authorizes probation, never commitment.

\section{Implementation}
\label{sec:impl}
The implementation comprises a Python controller and evidence pipeline,
generated C++ ABI code, an AArch64 Linux executor, and Arduino/Zephyr sketches.
The Linux executor is a persistent process that receives requests over a pipe
and reports its own compute time with a monotonic clock. The MCU executor
answers each request in a single Bridge call that returns both the result and
the compute time measured with the Zephyr cycle counter, so no diagnostic
calls fall on the measured path. Requests carry a 64-byte input block encoded
in base64 (at most 96 characters), inside the 128-byte payload envelope that
Section~\ref{sec:interconnect} verifies; both executors share one decoder so
that they perform identical work. Every result is checked against an
independent Python reference.

Instrumentation is layered. Reduced mode records core timing, full mode adds
structured observations, and disabled mode measures the base call path. An
optional kernel probe is evaluated separately and is not required for
control. An LED-matrix status display has disabled, static, and 10/30/60-Hz
refresh profiles so that its cost is measured rather than assumed. All
records carry schema versions, run identifiers, monotonic timestamps,
outcomes, and mode metadata, and command-line audits cross-check source
hashes, sample counts, derived deadlines, and reported numbers.

\section{Evaluation Methodology}
\label{sec:method}
The evaluation asks whether the protocol of Section~\ref{sec:design} ever has
a beneficial move to make on real hardware, and what it does when it does
not. This section describes how we measured both domains under matched
conditions so that the answer is not an artifact of the measurement.

\subsection{Testbed and workloads}
Table~\ref{tab:testbed} summarizes the setup. All measurements use one UNO Q with its stock Bridge stack.
The five workloads span a trivial task and a compute sweep: a signed 32-bit
addition; a bitwise CRC-32 over the 64-byte block; and a 16-tap triangular
Q15 FIR filter over 32 samples repeated 1, 64, or 512 times with a
pass-dependent perturbation, written FIR$\times$1, FIR$\times$64, and
FIR$\times$512.

\begin{table}[t]
\caption{Experimental setup}
\label{tab:testbed}
\centering
\small
\begin{tabular}{ll}
\toprule
Item & Configuration \\
\midrule
Board & Arduino UNO Q \\
Linux domain & Cortex-A53, Debian, AArch64 \\
MCU domain & STM32U585 Cortex-M33, Zephyr \\
Transport & Bridge RPC / MessagePack \\
Workloads & add, CRC-32, FIR$\times\{1,64,512\}$ \\
Samples per run & \StageTwelveIterations{} after \StageTwelveWarmup{} warm-up \\
Load conditions & idle, CPU, RT contention \\
Caller policies & \texttt{SCHED\_OTHER}, \texttt{SCHED\_FIFO} 50 \\
Per-task deadline & \StageTwelveMinNodeDeadline--\StageTwelveMaxNodeDeadline{} ms, Eq.~\eqref{eq:deadline} \\
Claim class & Empirical soft real-time \\
\bottomrule
\end{tabular}
\end{table}

\subsection{Load conditions and callers}
Three conditions are applied for the whole duration of a run. \emph{Idle}
applies no load. \emph{CPU load} runs eight \texttt{stress-ng} CPU workers
under \texttt{SCHED\_OTHER}. \emph{RT contention} runs four \texttt{stress-ng}
workers at \texttt{SCHED\_FIFO} priority 80 with a 50\% duty cycle in 10-ms
slices, modeling a co-located higher-priority real-time workload that tuning
the measured task cannot preempt. Under each condition both domains are
profiled with an ordinary caller and with the caller and Linux executor at
\texttt{SCHED\_FIFO} priority 50, giving 12 profile runs and 120{,}000 retained
invocations.

\subsection{Evidence classes}
We keep three evidence classes separate. \emph{Direct measurements} are the
per-call latencies above. \emph{Hardware scenarios} exercise the real
controller on the board and demonstrate one committed switch and one
rollback. \emph{Ordered trace replay} composes the measured traces into a
four-node chain: each task visit consumes the next sample of the relevant
trace in measured order, which preserves autocorrelation and regime changes,
and \StageTwelveChainObservations{} chain observations are replayed per cell.
Replay runs the protocol of Section~\ref{sec:design} exactly, including
Eqs.~\eqref{eq:deadline}--\eqref{eq:commit}. It permits paired comparison
across policies but is not equivalent to running a complete application.

\subsection{Baselines}
Table~\ref{tab:baselines} lists the policies. The oracle chooses, for every
task visit, the faster of the two domains' next measured samples; it is a
non-causal lower bound, not a competitor.

\begin{table}[t]
\caption{Policies compared on the four-node chain}
\label{tab:baselines}
\centering
\small
\begin{tabular}{lp{4.3cm}}
\toprule
Policy & Role \\
\midrule
Ordinary Linux & All tasks in Linux, default scheduling \\
Tuned Linux & All tasks in Linux, \texttt{SCHED\_FIFO} caller \\
MCU-first & All eligible tasks on the MCU \\
Manual expert & Declared placement L, M, M, L \\
\system static & Domain with lower idle p95 \\
\system recovery & Guarded recovery, ordinary caller \\
\system recovery (tuned) & Guarded recovery, \texttt{SCHED\_FIFO} caller \\
Offline oracle & Non-causal per-visit minimum \\
\bottomrule
\end{tabular}
\end{table}

\subsection{Statistics}
Primary metrics are p95 latency and deadline-miss rate. For per-call p95 we
report 95\% confidence intervals from a circular block bootstrap
(\StageTwelveBootstrapResamples{} resamples, block length
\StageTwelveBootstrapBlock), which respects the short-range autocorrelation of
measured traces. All tables, figures, and numbers in the text are generated
from the raw traces by scripts in the artifact, which record input digests.

\subsection{Relation to the earlier release}
\label{sec:correction}
An earlier artifact release~\cite{qfabric2026v101} included a manuscript
whose flagship result was a 36.6\% p95 reduction from recovery on a
replayed chain under Linux load. Re-examination found two defects. First,
its Linux executor was a new process per invocation, so Linux latency was
dominated by process creation: 8.05~ms p95 for an addition, against
\StageTwelveIdleLinuxAddP~ms with the persistent executor used here. Second,
the replay drew MCU latencies from idle traces while drawing Linux latencies
from loaded traces, although every MCU call is dispatched through Linux and
is affected by the same load. The present evaluation removes both defects
and fixes its deadline rule and replay protocol in the repository before any
new data was collected. Under the corrected method the earlier improvement
does not hold. The earlier release remains available unchanged for
transparency.

\section{Results}
\label{sec:results}
We answer the research questions in the order of the argument: first what it
costs to cross the boundary (RQ1), then whether migration can recover that
cost per call (RQ2) or per chain (RQ3), and finally whether the guards matter
(RQ4) and what they cost (RQ5).

\subsection{RQ1: Interconnect characterization}
\label{sec:interconnect}
All cross-domain costs in \system originate from the Bridge RPC path between
Linux and the MCU, so we characterized it before deriving any contract.
Each point is the mean of three repetition-level percentiles over 1,000
Linux-observed round trips, recomputed from \IcRawSamplesVerified{} raw samples
with zero failures. Under the default \texttt{schedutil} governor, an idle
zero-byte echo has a p50 of \IcZeroByteIdlePfifty{} ms and p95 of
\IcZeroByteIdlePninetyfive{} ms, sustaining \IcZeroByteIdleCallsPerSec{}
sequential calls/s; at the largest demonstrated-safe payload,
\IcMaxPayloadBytes{} bytes, p95 rises to \IcMaxPayloadIdlePninetyfive{} ms and
the rate falls to \IcMaxPayloadIdleCallsPerSec{} calls/s
(Fig.~\ref{fig:interconnect}a); latency grows nearly linearly with payload
size. A 256-byte probe caused sustained loss of MCU RPC service, so we
make no claim beyond \IcMaxPayloadBytes{} bytes.

Four \texttt{stress-ng} CPU workers lowered several medians but raised p95 by
\IcSchedutilLoadPninetyfiveMin{}--\IcSchedutilLoadPninetyfiveMax{} across
Linux-initiated calls. Pinning the \texttt{performance} governor did not remove
this tail penalty (\IcPerformanceLoadPninetyfiveMin{}--\IcPerformanceLoadPninetyfiveMax{});
it changed echo p95 by \IcGovernorIdlePninetyfiveMin{} to
\IcGovernorIdlePninetyfiveMax{} at idle and by \IcGovernorLoadedPninetyfiveMin{}
to \IcGovernorLoadedPninetyfiveMax{} under load. Frequency policy therefore does
not explain the loaded tail. With concurrent 8-byte callers, aggregate
throughput grows sublinearly from \IcConcOneIdleCallsPerSec{} to
\IcConcEightIdleCallsPerSec{} calls/s (\IcConcScaling$\times$) at one to eight
workers, while p95 rises from \IcConcOneIdlePninetyfive{} to
\IcConcEightIdlePninetyfive{} ms (Fig.~\ref{fig:interconnect}b). Load increased
throughput by \IcConcLoadedGainMin{}--\IcConcLoadedGainMax{} yet produced a
\IcConcLoadedMaxMs-ms outlier, so contracts must be derived from tails, not
throughput or medians.

\begin{figure*}[t]
\centering
\pgfplotstableread{generated/interconnect-payload.dat}\icpayload
\pgfplotstableread{generated/interconnect-concurrency.dat}\icconc
\begin{minipage}{0.49\textwidth}
\centering
\begin{tikzpicture}
\begin{axis}[
  width=\linewidth, height=48mm,
  xlabel={Payload (bytes)}, ylabel={Round-trip latency (ms)}, ymin=0,
  symbolic x coords={0,1,8,32,128}, xtick=data,
  grid=major, enlarge x limits=0.06,
  legend style={font=\scriptsize, at={(0.02,0.98)}, anchor=north west},
  legend cell align=left,
]
\addplot[qblue, thick, mark=*, mark size=1.4pt]
  table[x=payload_bytes, y=idle_p95_ms]{\icpayload};
\addlegendentry{idle p95}
\addplot[qorange, thick, mark=square*, mark size=1.4pt]
  table[x=payload_bytes, y=loaded_p95_ms]{\icpayload};
\addlegendentry{loaded p95}
\addplot[qblue, dashed, mark=o, mark size=1.4pt, mark options={solid}]
  table[x=payload_bytes, y=idle_p50_ms]{\icpayload};
\addlegendentry{idle p50}
\addplot[qorange, dashed, mark=square, mark size=1.4pt, mark options={solid}]
  table[x=payload_bytes, y=loaded_p50_ms]{\icpayload};
\addlegendentry{loaded p50}
\end{axis}
\end{tikzpicture}

\small (a) Payload sweep, one sequential caller
\end{minipage}
\begin{minipage}{0.49\textwidth}
\centering
\begin{tikzpicture}
\begin{axis}[
  width=\linewidth, height=48mm,
  xlabel={Aggregate throughput (calls/s)}, ylabel={p95 latency (ms)},
  ymin=0, grid=major,
  legend style={font=\scriptsize, at={(0.02,0.98)}, anchor=north west},
  legend cell align=left,
  nodes near coords, point meta=explicit symbolic,
  nodes near coords style={font=\scriptsize, anchor=south east},
]
\addplot[qblue, thick, mark=*, mark size=1.4pt]
  table[x=idle_calls_per_s, y=idle_p95_ms, meta=workers]{\icconc};
\addlegendentry{idle}
\addplot[qorange, thick, mark=square*, mark size=1.4pt,
  nodes near coords style={anchor=north west}]
  table[x=loaded_calls_per_s, y=loaded_p95_ms, meta=workers]{\icconc};
\addlegendentry{loaded}
\end{axis}
\end{tikzpicture}

\small (b) 1/2/4/8 concurrent workers, 8-byte payload
\end{minipage}
\caption{Linux$\leftrightarrow$MCU Bridge RPC characterization under the
\texttt{schedutil} governor. Values are means of three repetition-level
percentiles; ``loaded'' runs four \texttt{stress-ng} CPU workers. Point labels in
(b) give the worker count.}
\label{fig:interconnect}
\end{figure*}
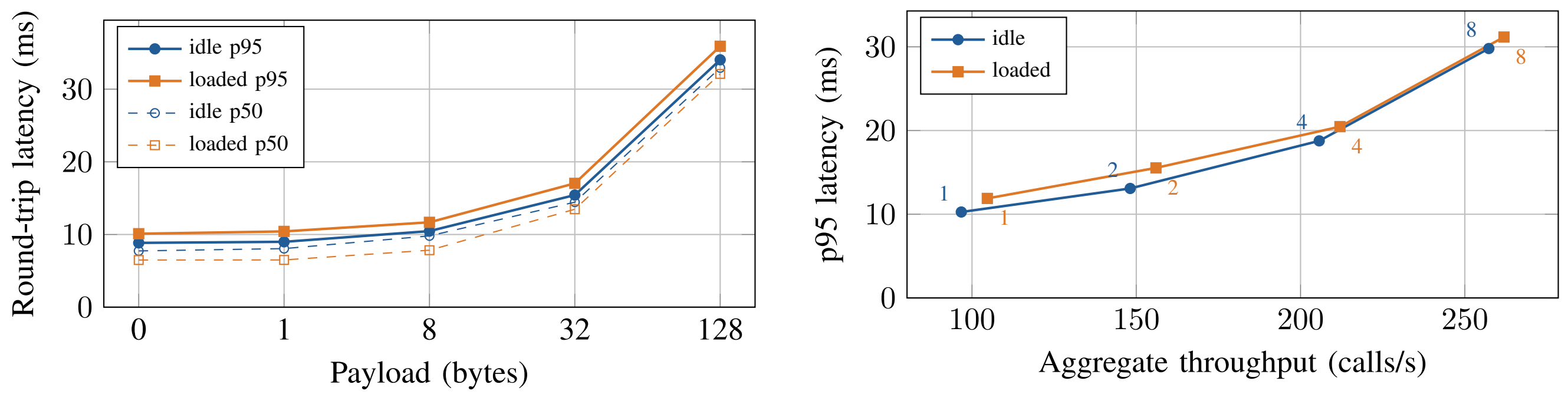

\begin{figure*}[t]
\centering
\pgfplotstableread{generated/stage12-percall.dat}\percalldata
\begin{tikzpicture}
\begin{axis}[ybar, width=0.36\textwidth, height=44mm, ymode=log, log origin=infty,     ymin=0.03, ymax=1000,     symbolic x coords={add,crc32,fir-x1,fir-x64,fir-x512},     xtick=data, xticklabels={add,CRC-32,FIR$\times$1,FIR$\times$64,FIR$\times$512},     x tick label style={rotate=30,anchor=east,font=\scriptsize},     bar width=5pt, enlarge x limits=0.14, grid=major,     legend style={font=\scriptsize,at={(0.03,0.97)},anchor=north west}, title={\small Idle}, ylabel={p95 per call (ms)}]
\addplot[fill=qgreen!70,draw=qgreen] table[x=workload,y=idle_linux]{\percalldata};
\addplot[fill=qorange!75,draw=qorange] table[x=workload,y=idle_rt]{\percalldata};
\legend{Linux,MCU}
\end{axis}
\end{tikzpicture}\hfill
\begin{tikzpicture}
\begin{axis}[ybar, width=0.36\textwidth, height=44mm, ymode=log, log origin=infty,     ymin=0.03, ymax=1000,     symbolic x coords={add,crc32,fir-x1,fir-x64,fir-x512},     xtick=data, xticklabels={add,CRC-32,FIR$\times$1,FIR$\times$64,FIR$\times$512},     x tick label style={rotate=30,anchor=east,font=\scriptsize},     bar width=5pt, enlarge x limits=0.14, grid=major,     legend style={font=\scriptsize,at={(0.03,0.97)},anchor=north west}, title={\small CPU load}, yticklabels={}]
\addplot[fill=qgreen!70,draw=qgreen] table[x=workload,y=cpu_linux]{\percalldata};
\addplot[fill=qorange!75,draw=qorange] table[x=workload,y=cpu_rt]{\percalldata};
\end{axis}
\end{tikzpicture}\hfill
\begin{tikzpicture}
\begin{axis}[ybar, width=0.36\textwidth, height=44mm, ymode=log, log origin=infty,     ymin=0.03, ymax=1000,     symbolic x coords={add,crc32,fir-x1,fir-x64,fir-x512},     xtick=data, xticklabels={add,CRC-32,FIR$\times$1,FIR$\times$64,FIR$\times$512},     x tick label style={rotate=30,anchor=east,font=\scriptsize},     bar width=5pt, enlarge x limits=0.14, grid=major,     legend style={font=\scriptsize,at={(0.03,0.97)},anchor=north west}, title={\small RT contention}, yticklabels={}]
\addplot[fill=qgreen!70,draw=qgreen] table[x=workload,y=rtcontention_linux]{\percalldata};
\addplot[fill=qorange!75,draw=qorange] table[x=workload,y=rtcontention_rt]{\percalldata};
\end{axis}
\end{tikzpicture}
\caption{Per-call p95 latency with ordinary callers, logarithmic scale. Linux
contention moves the MCU path more than the Linux path because each MCU call
is issued and relayed by Linux processes.}
\label{fig:percall}
\end{figure*}
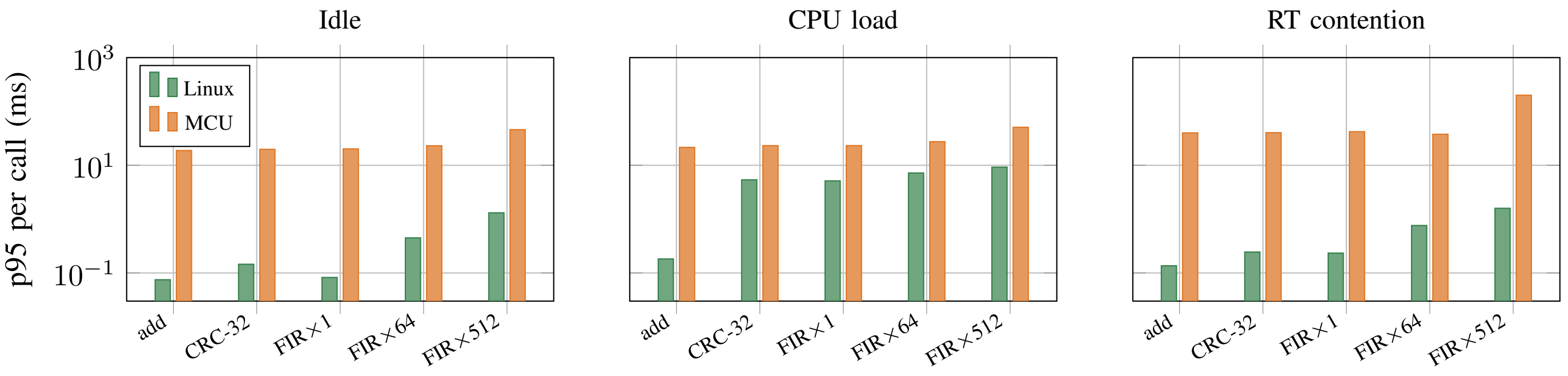

\subsection{RQ2: Matched-load per-call comparison}
\label{sec:percall}
Table~\ref{tab:percall} and Fig.~\ref{fig:percall} compare the domains per
call. At idle, an MCU addition takes \StageTwelveIdleRtAddP~ms at p95 against
\StageTwelveIdleLinuxAddP~ms in Linux, a \StageTwelveIdleRatio$\times$ gap.
Table~\ref{tab:decomposition} decomposes the idle median with
Eq.~\eqref{eq:decomposition}. The MCU residual stays near 19~ms for every
workload, so dispatch, not compute, dominates the MCU path; and the MCU
computes FIR$\times$512 \StageTwelveFirComputeRatio$\times$ more slowly than
the Cortex-A53, so heavier kernels widen the gap rather than closing it.

\begin{table}[!tb]
\caption{Per-call p95 latency (ms) with ordinary callers. Brackets give 95\%
block-bootstrap confidence intervals.}
\label{tab:percall}
\centering
\scriptsize
\setlength{\tabcolsep}{4pt}
\input{generated/stage12-percall-table.tex}
\end{table}

\begin{table}[!tb]
\caption{Idle per-call decomposition (Eq.~\ref{eq:decomposition}). Compute is
measured inside each executor; the residual is the difference of medians.}
\label{tab:decomposition}
\centering
\footnotesize
\setlength{\tabcolsep}{4pt}
\input{generated/stage12-decomposition-table.tex}
\end{table}

Load changes the picture in an instructive way. CPU load raises the
ordinary Linux p95 to a few milliseconds, which a \texttt{SCHED\_FIFO} caller
removes almost entirely (Table~\ref{tab:chain}, tuned Linux). Under RT contention, the Linux path keeps a p95
below 2~ms with occasional stalls, while the MCU path degrades sharply, with
p99 up to \StageTwelveContentionRtPNinetyNine~ms against
\StageTwelveContentionLinuxPNinetyNine~ms for Linux. An MCU invocation
traverses the caller, the Bridge router, and serial-driver activity, all on
Linux, so it is exposed to the co-runner at several points, whereas a Linux
invocation needs one runnable core at one point. Moving a task to the MCU
does not move its dispatch path off Linux.

\subsection{RQ3: Chain-level outcomes and controller decisions}
\label{sec:chain}
Table~\ref{tab:chain} reports the replayed four-node chain. Linux meets the
contract in every cell: the worst ordinary-Linux p95 is
\StageTwelveWorstLinuxChainP~ms against graph deadlines of at least
\the\numexpr4*\StageTwelveMinNodeDeadline\relax~ms, and the worst miss rate
of either Linux policy is \StageTwelveWorstLinuxMiss. MCU-first is several
times slower in every cell. The offline oracle coincides with Linux except
for rare visits in which a stalled Linux sample exceeds the paired MCU sample.
With the hysteresis of Eq.~\eqref{eq:hysteresis}, \system performs
\StageTwelveSwitches{} switches across all \StageTwelveCells{} cells, which is
the correct decision: no admissible alternative would have improved the
contract.

\begin{table}[!tb]
\caption{Replayed four-node chain, \StageTwelveChainObservations{}
observations per cell; latencies in ms. $D_G$ is the graph deadline of
Eq.~\eqref{eq:deadline}. Neither \system recovery variant switched in any
cell.}
\label{tab:chain}
\centering
\scriptsize
\setlength{\tabcolsep}{2.6pt}
\input{generated/stage12-chain-table.tex}
\end{table}

\subsection{RQ4: Are the guards necessary?}
\label{sec:guards}
Two kinds of evidence address the guards. On hardware, a controlled scenario
detected a sustained violation one complete window after onset with no false
switch, switched to a feasible alternate, and verified it after four windows;
a separate injected scenario forced probation failure and produced a
rollback. A negative suite exercised nine stable rejection reasons:
insufficient evidence, unsatisfied contract, semantic ineligibility,
destination admission failure, MCU headroom, predicted deadline miss, higher
predicted end-to-end cost, observed miss-rate failure, and active cooldown.
These scenarios demonstrate the protocol's mechanics; their latencies are
not used as performance evidence.

Table~\ref{tab:ablation} isolates hysteresis on the Stage 12 traces by
setting $k=1$ in Eq.~\eqref{eq:hysteresis}. The ablated controller reacts to
isolated Linux stalls under RT contention and switches in
\StageTwelveAblatedSwitches{} cells, raising p95 by up to
\StageTwelveAblationWorstRatio$\times$ and increasing miss rates. Probation,
Eq.~\eqref{eq:commit}, rolls back \StageTwelveAblatedRollbacks{} of these
moves; the remaining two were still in probation when the trace ended (``open at end'').
Hysteresis prevents the harmful moves, and probation bounds the damage when
a move is made anyway.

\begin{table}[!tb]
\caption{Hysteresis ablation under RT contention: switching on the first
violating window ($k=1$) instead of $k=3$.}
\label{tab:ablation}
\centering
\scriptsize
\setlength{\tabcolsep}{3pt}
\input{generated/stage12-ablation-table.tex}
\end{table}

\subsection{RQ5: Mechanism overhead}
The mechanism itself is inexpensive. One policy
decision costs \PolicyCpuP\,$\mu$s of CPU time and \PolicyWallP\,$\mu$s of wall
time at p95 over 1,000 in-process evaluations, with a \PolicyPeakBytes-byte
peak observed by \texttt{tracemalloc}; the latter excludes the interpreter
baseline and is not total process memory. A decision is thus two orders of
magnitude cheaper than one MCU dispatch (\StageTwelveIdleRtAddP~ms at p95 for
an idle addition). The worst LED-matrix refresh profile
changes Bridge p95 by \VisualizationOverhead.

\begin{table}[!tb]
\caption{Answers to the research questions}
\label{tab:rq}
\centering
\small
\begin{tabular}{lp{6.0cm}}
\toprule
RQ & Answer \\
\midrule
RQ1 & About 9~ms p95 per zero-byte call, rising with payload; tails grow under load regardless of governor. \\
RQ2 & Linux is faster in every cell; MCU dispatch is a fixed $\approx$19~ms and Linux contention hurts the MCU path more. \\
RQ3 & No contract fails on Linux; \system correctly makes \StageTwelveSwitches{} switches in \StageTwelveCells{} cells. \\
RQ4 & Yes: without hysteresis, \StageTwelveAblatedSwitches{} harmful switches; probation rolls back \StageTwelveAblatedRollbacks. \\
RQ5 & A decision costs about \PolicyWallP\,$\mu$s, negligible against dispatch. \\
\bottomrule
\end{tabular}
\end{table}

\section{Discussion}
\label{sec:discussion}
\subsection{When can cross-domain recovery pay off?}
Equation~\eqref{eq:decomposition} makes the boundary explicit. Migrating task
$v$ from Linux to the MCU helps only if
$C_{v,\mathrm{M}}+R_{v,\mathrm{M}} < C_{v,\mathrm{L}}+R_{v,\mathrm{L}}$ under
the degraded condition. On this board $R_{v,\mathrm{M}}\approx19$~ms at idle,
$C_{v,\mathrm{M}}$ exceeds $C_{v,\mathrm{L}}$ by more than an order of
magnitude, and the interference that inflates $R_{v,\mathrm{L}}$ inflates
$R_{v,\mathrm{M}}$ at least as much, because both residuals are paid in Linux
userspace. The inequality therefore fails for any task dispatched one call at
a time from Linux. Three architectural changes could reverse it: a
lower-latency transport such as shared-memory signaling, which reaches
sub-microsecond latency on other heterogeneous SoCs~\cite{coppik2025lowlatency};
a dispatcher outside Linux userspace, for example a kernel driver or a
partitioned core~\cite{martins2023partitioning}; or MCU-resident chains whose
release is triggered by an MCU timer so that only inputs and results cross
the boundary. Each is a direct test of the boundary reported here.

\subsection{What the guards contribute}
Recovery and tuning solve different problems. When the workload can remain in
Linux, a \texttt{SCHED\_FIFO} caller removes most CPU-load interference and is
the right first step. The value of \system's guards is that they make
adaptive placement safe to leave enabled: hysteresis ignores isolated stalls,
admission rejects moves predicted to miss, and probation reverses moves whose
benefit does not materialize. Table~\ref{tab:ablation} shows that a plausible
but unguarded controller would have turned sub-millisecond tails into
100-ms tails.

\subsection{Programmer effort}
A task declaration states types, effect, and identifier once, and the
generated ABI removes hand-written marshaling. The programmer must still
supply semantically equivalent implementations for both domains, choose a
defensible contract, and state Linux-only effects; \system shifts effort from
transport plumbing to contracts and equivalence rather than removing it.

\subsection{Threats to validity}
\label{sec:threats}
\textbf{External validity.} Results come from one UNO Q, one software stack,
and five small kernels. Boards with faster transports or dispatch outside
Linux userspace may reverse the inequality above. The chain results replay
measured traces rather than running a full application.

\textbf{Construct validity.} Percentiles and miss rates do not establish
worst-case execution time or hard real-time guarantees, and the admission
test of Eq.~\eqref{eq:admission} is a utilization gate, not a schedulability
proof. The residual in Table~\ref{tab:decomposition} is a difference of
medians, not a measured component.

\textbf{Internal validity.} All runs share one board and may be affected by
firmware, thermal state, and background activity. Ordered replay preserves
temporal structure within a trace but cannot represent interactions between
tasks executing concurrently. The deadline rule and replay protocol were
committed to the repository before any Stage 12 data was collected.

\textbf{Baseline and load validity.} \texttt{SCHED\_FIFO} is a strong but not
exhaustive Linux baseline; \texttt{SCHED\_DEADLINE}~\cite{lelli2016deadline},
CPU isolation, and affinity warrant comparison. The RT-contention co-runner is
one synthetic pattern; synchronized or fully saturating interference would
change magnitudes, though it would burden the MCU dispatch path as well.

\textbf{Safety scope.} Quiescent switching prevents mixed-epoch calls in the
prototype and effect annotations reject known-ineligible tasks, but this is
not a proof of implementation equivalence. Production use would need state
transfer, authentication, fault containment, and multi-client analysis.

\FloatBarrier
\section{Conclusion}
\label{sec:conclusion}
\system demonstrates a guarded form of runtime adaptation across a
Linux--RTOS boundary: an empirical contract violation can trigger a
quiescent, admitted switch whose benefit must be observed before it is kept.
Measured on an Arduino UNO Q with matched load conditions, the Linux--MCU
path costs about 19~ms per call, the MCU computes up to
\StageTwelveFirComputeRatio$\times$ more slowly, and Linux interference
reaches the MCU path through its Linux-resident dispatch. Migration
consequently never improves a contract in our study, the controller correctly
declines to migrate, and an unguarded controller would have made harmful
moves that only probation undoes. The broader lesson for heterogeneous
embedded architectures is that the isolation offered by a real-time core is
available only to work whose dispatch also leaves the general-purpose
domain. Testing faster transports, non-Linux dispatch, and MCU-resident
chains is the natural next step.

\section*{Artifact Availability}
The repository~\cite{qfabric2026} contains source, generated code, unit
tests, board sketches, raw and processed traces, campaign logs, audits, and
the generators for every number, table, and figure in this paper; running
\texttt{make} in the \texttt{paper} directory regenerates the assets and
builds the PDF. The release cited here, \texttt{v1.1.0}, is archived on Zenodo
with DOI \href{https://doi.org/10.5281/zenodo.23692354}{\texttt{10.5281/zenodo.23692354}}.

\bibliographystyle{IEEEtran}
\bibliography{references}

\end{document}

%% file: generated/metrics.tex
\newcommand{\PolicyCpuP}{170.1}
\newcommand{\PolicyWallP}{177.2}
\newcommand{\PolicyPeakBytes}{3138}

\newcommand{\VisualizationOverhead}{2.05\%}

%% file: generated/interconnect-metrics.tex
\newcommand{\IcMaxPayloadBytes}{128}
\newcommand{\IcZeroByteIdleCallsPerSec}{127.2}

\newcommand{\IcMaxPayloadIdleCallsPerSec}{30.2}

\newcommand{\IcZeroByteIdlePfifty}{7.74}
\newcommand{\IcZeroByteIdlePninetyfive}{8.85}

\newcommand{\IcMaxPayloadIdlePninetyfive}{34.04}

\newcommand{\IcSchedutilLoadPninetyfiveMin}{5.5\%}
\newcommand{\IcSchedutilLoadPninetyfiveMax}{23.5\%}
\newcommand{\IcPerformanceLoadPninetyfiveMin}{6.7\%}
\newcommand{\IcPerformanceLoadPninetyfiveMax}{42.9\%}
\newcommand{\IcGovernorIdlePninetyfiveMin}{$-11.1$\%}
\newcommand{\IcGovernorIdlePninetyfiveMax}{$-1.3$\%}
\newcommand{\IcGovernorLoadedPninetyfiveMin}{$-0.2$\%}
\newcommand{\IcGovernorLoadedPninetyfiveMax}{$+10.6$\%}
\newcommand{\IcConcOneIdleCallsPerSec}{96.75}
\newcommand{\IcConcEightIdleCallsPerSec}{257.31}

\newcommand{\IcConcScaling}{2.66}
\newcommand{\IcConcOneIdlePninetyfive}{10.27}
\newcommand{\IcConcEightIdlePninetyfive}{29.80}

\newcommand{\IcConcLoadedGainMin}{1.8\%}
\newcommand{\IcConcLoadedGainMax}{8.2\%}
\newcommand{\IcConcLoadedMaxMs}{115.82}
\newcommand{\IcRawSamplesVerified}{108{,}000}

%% file: generated/stage12-metrics.tex
\newcommand{\StageTwelveIterations}{2000}
\newcommand{\StageTwelveWarmup}{100}
\newcommand{\StageTwelveChainObservations}{500}
\newcommand{\StageTwelveCells}{30}
\newcommand{\StageTwelveSwitches}{0}
\newcommand{\StageTwelveAblatedSwitches}{7}
\newcommand{\StageTwelveAblatedRollbacks}{5}
\newcommand{\StageTwelveAblationWorstRatio}{177}
\newcommand{\StageTwelveIdleRtAddP}{18.81}
\newcommand{\StageTwelveIdleLinuxAddP}{0.075}
\newcommand{\StageTwelveIdleRatio}{251}
\newcommand{\StageTwelveFirComputeRatio}{34}
\newcommand{\StageTwelveContentionRtPNinetyNine}{400}
\newcommand{\StageTwelveContentionLinuxPNinetyNine}{17.0}
\newcommand{\StageTwelveWorstLinuxChainP}{26.7}
\newcommand{\StageTwelveWorstLinuxMiss}{0.8\%}
\newcommand{\StageTwelveMinNodeDeadline}{24}
\newcommand{\StageTwelveMaxNodeDeadline}{58}
\newcommand{\StageTwelveBootstrapResamples}{2,000}
\newcommand{\StageTwelveBootstrapBlock}{50}

%% file: generated/stage12-percall-table.tex
\begin{tabular}{llrr}
\toprule
Load & Workload & Linux p95 [95\% CI] & MCU p95 [95\% CI] \\
\midrule
Idle & add & 0.07 [0.07, 0.08] & 18.81 [18.79, 18.84] \\
 & CRC-32 & 0.15 [0.12, 0.16] & 19.78 [19.76, 19.79] \\
 & FIR$\times$1 & 0.08 [0.08, 0.09] & 20.16 [20.13, 20.19] \\
 & FIR$\times$64 & 0.45 [0.44, 0.46] & 23.07 [23.00, 23.14] \\
 & FIR$\times$512 & 1.31 [1.22, 1.33] & 45.93 [45.88, 46.02] \\
\midrule
CPU & add & 0.18 [0.14, 0.23] & 21.49 [21.16, 21.86] \\
 & CRC-32 & 5.37 [5.22, 5.55] & 23.17 [22.72, 23.35] \\
 & FIR$\times$1 & 5.13 [4.94, 5.41] & 23.22 [22.80, 23.51] \\
 & FIR$\times$64 & 7.20 [6.56, 7.46] & 27.42 [27.27, 27.75] \\
 & FIR$\times$512 & 9.24 [9.11, 9.39] & 50.76 [50.31, 51.09] \\
\midrule
RT & add & 0.14 [0.10, 0.17] & 39.99 [38.74, 42.70] \\
 & CRC-32 & 0.24 [0.19, 0.27] & 40.33 [38.48, 47.21] \\
 & FIR$\times$1 & 0.23 [0.17, 0.26] & 42.04 [39.24, 44.47] \\
 & FIR$\times$64 & 0.76 [0.74, 0.78] & 37.70 [36.36, 39.10] \\
 & FIR$\times$512 & 1.59 [1.54, 1.67] & 200.37 [195.34, 219.40] \\
\bottomrule
\end{tabular}

%% file: generated/stage12-decomposition-table.tex
\begin{tabular}{lrrrr}
\toprule
 & \multicolumn{2}{c}{Compute p50 ($\mu$s)} & \multicolumn{2}{c}{Residual p50 (ms)} \\
\cmidrule(lr){2-3}\cmidrule(lr){4-5}
Workload & Linux & MCU & Linux & MCU \\
\midrule
add & 0.1 & 0.8 & 0.053 & 18.53 \\
CRC-32 & 1.3 & 26.9 & 0.060 & 19.09 \\
FIR$\times$1 & 1.7 & 52.4 & 0.060 & 19.27 \\
FIR$\times$64 & 95.7 & 3231.2 & 0.277 & 18.85 \\
FIR$\times$512 & 760.4 & 25842.1 & 0.380 & 19.12 \\
\bottomrule
\end{tabular}

%% file: generated/stage12-chain-table.tex
\begin{tabular}{llrrrrrr}
\toprule
 & & & \multicolumn{2}{c}{Ordinary Linux} & \multicolumn{2}{c}{Tuned Linux} & MCU-first \\
\cmidrule(lr){4-5}\cmidrule(lr){6-7}\cmidrule(lr){8-8}
Load & Workload & $D_G$ & p95 & miss (\%) & p95 & miss (\%) & p95 \\
\midrule
Idle & add & 96 & 0.27 & 0.0 & 0.22 & 0.0 & 74.6 \\
 & CRC-32 & 100 & 0.52 & 0.0 & 0.26 & 0.0 & 78.5 \\
 & FIR$\times$1 & 104 & 0.32 & 0.0 & 0.25 & 0.0 & 78.6 \\
 & FIR$\times$64 & 120 & 1.69 & 0.0 & 1.63 & 0.0 & 90.5 \\
 & FIR$\times$512 & 232 & 4.99 & 0.0 & 4.65 & 0.0 & 181.9 \\
\midrule
CPU & add & 96 & 5.34 & 0.0 & 0.40 & 0.0 & 76.5 \\
 & CRC-32 & 100 & 9.34 & 0.0 & 0.64 & 0.0 & 83.0 \\
 & FIR$\times$1 & 104 & 8.81 & 0.0 & 0.66 & 0.0 & 82.0 \\
 & FIR$\times$64 & 120 & 17.31 & 0.0 & 1.69 & 0.0 & 101.3 \\
 & FIR$\times$512 & 232 & 26.70 & 0.0 & 4.52 & 0.0 & 191.3 \\
\midrule
RT & add & 96 & 0.58 & 0.2 & 0.31 & 0.0 & 444.5 \\
 & CRC-32 & 100 & 0.85 & 0.2 & 0.56 & 0.0 & 458.9 \\
 & FIR$\times$1 & 104 & 0.81 & 0.2 & 0.71 & 0.2 & 452.8 \\
 & FIR$\times$64 & 120 & 3.97 & 0.8 & 2.97 & 0.2 & 253.6 \\
 & FIR$\times$512 & 232 & 20.07 & 0.0 & 17.00 & 0.8 & 539.3 \\
\bottomrule
\end{tabular}

%% file: generated/stage12-ablation-table.tex
\begin{tabular}{llrrrl}
\toprule
Workload & Caller & Win. & p95 (ms) & Miss (\%) & Outcome \\
\midrule
add & ordinary & 16 & 0.58$\rightarrow$103.32 & 0.2$\rightarrow$6.4 & rolled back \\
CRC-32 & ordinary & 21 & 0.85$\rightarrow$95.05 & 0.2$\rightarrow$3.8 & open at end \\
FIR$\times$1 & ordinary & 15 & 0.81$\rightarrow$102.03 & 0.2$\rightarrow$4.8 & rolled back \\
FIR$\times$1 & FIFO & 7 & 0.71$\rightarrow$102.89 & 0.2$\rightarrow$4.4 & rolled back \\
FIR$\times$64 & ordinary & 4 & 3.97$\rightarrow$109.51 & 0.8$\rightarrow$3.8 & rolled back \\
FIR$\times$64 & FIFO & 21 & 2.97$\rightarrow$101.46 & 0.2$\rightarrow$0.8 & open at end \\
FIR$\times$512 & FIFO & 9 & 17.00$\rightarrow$379.48 & 0.8$\rightarrow$10.4 & rolled back \\
\bottomrule
\end{tabular}

%% file: qfabric-paper.bbl
\begin{thebibliography}{10}
\providecommand{\url}[1]{#1}
\csname url@samestyle\endcsname
\providecommand{\newblock}{\relax}
\providecommand{\bibinfo}[2]{#2}
\providecommand{\BIBentrySTDinterwordspacing}{\spaceskip=0pt\relax}
\providecommand{\BIBentryALTinterwordstretchfactor}{4}
\providecommand{\BIBentryALTinterwordspacing}{\spaceskip=\fontdimen2\font plus
\BIBentryALTinterwordstretchfactor\fontdimen3\font minus \fontdimen4\font\relax}
\providecommand{\BIBforeignlanguage}[2]{{%
\expandafter\ifx\csname l@#1\endcsname\relax
\typeout{** WARNING: IEEEtran.bst: No hyphenation pattern has been}%
\typeout{** loaded for the language `#1'. Using the pattern for}%
\typeout{** the default language instead.}%
\else
\language=\csname l@#1\endcsname
\fi
#2}}
\providecommand{\BIBdecl}{\relax}
\BIBdecl

\bibitem{arduinoUnoQ}
\BIBentryALTinterwordspacing
{Arduino}, \emph{{UNO Q} User Manual}, Arduino, 2026, accessed 2026-10-09. [Online]. Available: \url{https://docs.arduino.cc/tutorials/uno-q/user-manual/}
\BIBentrySTDinterwordspacing

\bibitem{openampProject}
\BIBentryALTinterwordspacing
{OpenAMP Project}, ``The {OpenAMP} project,'' Hosted by Linaro, 2026, accessed 2026-10-27. [Online]. Available: \url{https://www.openampproject.org/}
\BIBentrySTDinterwordspacing

\bibitem{topcuoglu2002heft}
H.~Topcuoglu, S.~Hariri, and M.-Y. Wu, ``Performance-effective and low-complexity task scheduling for heterogeneous computing,'' \emph{IEEE Transactions on Parallel and Distributed Systems}, vol.~13, no.~3, pp. 260--274, 2002.

\bibitem{lu2002feedback}
C.~Lu, J.~A. Stankovic, G.~Tao, and S.~H. Son, ``Feedback control real-time scheduling: Framework, modeling, and algorithms,'' \emph{Real-Time Systems}, vol.~23, pp. 85--126, 2002.

\bibitem{reinbacher2014runtime}
T.~Reinbacher, M.~F\"ugger, and J.~Brauer, ``Runtime verification of embedded real-time systems,'' \emph{Formal Methods in System Design}, vol.~44, pp. 203--239, 2014.

\bibitem{nandi2015contracts}
P.~R. Nandi, A.~Monot, and M.~Oriol, ``Stochastic contracts for runtime checking of component-based real-time systems,'' in \emph{Proceedings of the 18th International ACM SIGSOFT Symposium on Component-Based Software Engineering}, 2015, pp. 138--147.

\bibitem{qfabric2026v101}
\BIBentryALTinterwordspacing
A.~Bhattacharya and A.~Chakraborty, ``{QFabric}: Contract-driven cross-domain task recovery,'' GitHub release v1.0.1, 2026, earlier artifact release; its manuscript is superseded by this paper. [Online]. Available: \url{https://github.com/elixpo/qfabric/releases/tag/v1.0.1}
\BIBentrySTDinterwordspacing

\bibitem{ramsauer2017jailhouse}
R.~Ramsauer, J.~Kiszka, D.~Lohmann, and W.~Mauerer, ``Look mum, no {VM} exits! (almost),'' in \emph{Proceedings of the 13th Workshop on Operating Systems Platforms for Embedded Real-Time Applications (OSPERT)}, 2017.

\bibitem{martins2020bao}
J.~Martins, A.~Tavares, M.~Solieri, M.~Bertogna, and S.~Pinto, ``{Bao}: A lightweight static partitioning hypervisor for modern multi-core embedded systems,'' in \emph{Next Generation Real-Time Embedded Systems}, ser. OpenAccess Series in Informatics, vol.~77, 2020, pp. 3:1--3:14.

\bibitem{real2004mode}
J.~Real and A.~Crespo, ``Mode change protocols for real-time systems: A survey and a new proposal,'' \emph{Real-Time Systems}, vol.~26, no.~2, pp. 161--197, 2004.

\bibitem{goraczko2006comos}
M.~Goraczko, C.-C. Han, J.~Helander, J.~Liu, B.~Priyantha, and F.~Zhao, ``{CoMOS}: An operating system for heterogeneous multi-processor sensor devices,'' Microsoft Research, Tech. Rep. MSR-TR-2006-177, 2006.

\bibitem{rouxel2021yasmin}
B.~Rouxel, S.~Altmeyer, and C.~Grelck, ``{YASMIN}: A real-time middleware for {COTS} heterogeneous platforms,'' in \emph{Proceedings of the 22nd International Middleware Conference}, 2021, pp. 298--309.

\bibitem{arduinoRouterBridge}
\BIBentryALTinterwordspacing
{Arduino}, \emph{{Arduino RouterBridge} Library}, Arduino, 2026, accessed 2026-10-09. [Online]. Available: \url{https://github.com/arduino-libraries/Arduino_RouterBridge}
\BIBentrySTDinterwordspacing

\bibitem{linuxRemoteproc}
\BIBentryALTinterwordspacing
{Linux Kernel Developers}, \emph{{Remote Processor Framework}}, The Linux Kernel Documentation, 2026, accessed 2026-10-09. [Online]. Available: \url{https://docs.kernel.org/staging/remoteproc.html}
\BIBentrySTDinterwordspacing

\bibitem{linuxRpmsg}
\BIBentryALTinterwordspacing
------, \emph{{Remote Processor Messaging (rpmsg) Framework}}, The Linux Kernel Documentation, 2024, accessed 2026-10-09. [Online]. Available: \url{https://www.kernel.org/doc/html/v6.8/staging/rpmsg.html}
\BIBentrySTDinterwordspacing

\bibitem{vestal2007preemptive}
S.~Vestal, ``Preemptive scheduling of multi-criticality systems with varying degrees of execution time assurance,'' in \emph{Proceedings of the 28th IEEE International Real-Time Systems Symposium (RTSS)}, 2007, pp. 239--243.

\bibitem{pourmohseni2020migration}
B.~Pourmohseni, A.~Smirnov, S.~Wildermann, and J.~Teich, ``Real-time task migration for dynamic resource management in many-core systems,'' in \emph{Next Generation Real-Time Embedded Systems}, ser. OpenAccess Series in Informatics, vol.~77, 2020, pp. 5:1--5:14.

\bibitem{yano2026awkernel}
A.~Yano, R.~Nishimura, Y.~Kobayashi, Y.~Takano, T.~Ishikawa-Aso, and T.~Azumi, ``{Awkernel}: {RTOS} bridging {DAG} scheduling theory and component-oriented real-time systems practice,'' 2026.

\bibitem{liu1973scheduling}
C.~L. Liu and J.~W. Layland, ``Scheduling algorithms for multiprogramming in a hard-real-time environment,'' \emph{Journal of the ACM}, vol.~20, no.~1, pp. 46--61, 1973.

\bibitem{kephart2003autonomic}
J.~O. Kephart and D.~M. Chess, ``The vision of autonomic computing,'' \emph{Computer}, vol.~36, no.~1, pp. 41--50, 2003.

\bibitem{alkhatib2016timing}
M.~A. Khatib, A.~Girard, and T.~Dang, ``Verification and synthesis of timing contracts for embedded controllers,'' in \emph{Proceedings of the 19th International Conference on Hybrid Systems: Computation and Control}, 2016.

\bibitem{camilli2018adaptive}
M.~Camilli, A.~Gargantini, and P.~Scandurra, ``Zone-based formal specification and timing analysis of real-time self-adaptive systems,'' \emph{Science of Computer Programming}, vol. 159, pp. 28--57, 2018.

\bibitem{alonso2021openamp}
S.~Alonso, J.~Lazaro, J.~Jimenez, L.~Muguira, and U.~Bidarte, ``Evaluating the {OpenAMP} framework in real-time embedded {SoC} platforms,'' in \emph{Proceedings of the 36th Conference on Design of Circuits and Integrated Systems (DCIS)}, 2021, pp. 1--6.

\bibitem{coppik2025lowlatency}
N.~Coppik, P.~Becker, and M.~Ritter, ``Low-latency real-time applications on heterogeneous {MPSoCs},'' in \emph{Next Generation Real-Time Embedded Systems}, ser. OpenAccess Series in Informatics, vol. 128, 2025, pp. 2:1--2:14.

\bibitem{martins2023partitioning}
J.~Martins and S.~Pinto, ``Shedding light on static partitioning hypervisors for {Arm}-based mixed-criticality systems,'' in \emph{Proceedings of the 29th IEEE Real-Time and Embedded Technology and Applications Symposium (RTAS)}, 2023, pp. 40--53.

\bibitem{gracioli2019mixed}
G.~Gracioli, R.~Tabish, R.~Mancuso, R.~Mirosanlou, R.~Pellizzoni, and M.~Caccamo, ``Designing mixed criticality applications on modern heterogeneous {MPSoC} platforms,'' in \emph{31st Euromicro Conference on Real-Time Systems (ECRTS 2019)}, ser. Leibniz International Proceedings in Informatics, vol. 133, 2019, pp. 27:1--27:25.

\bibitem{wilhelm2008wcet}
R.~Wilhelm, J.~Engblom, A.~Ermedahl, N.~Holsti, S.~Thesing, D.~Whalley, G.~Bernat, C.~Ferdinand, R.~Heckmann, T.~Mitra, F.~Mueller, I.~Puaut, P.~Puschner, J.~Staschulat, and P.~Stenstr\"om, ``The worst-case execution-time problem---overview of methods and survey of tools,'' \emph{ACM Transactions on Embedded Computing Systems}, vol.~7, no.~3, pp. 1--53, 2008.

\bibitem{cucugrosjean2012mbpta}
L.~Cucu-Grosjean, L.~Santinelli, M.~Houston, C.~Lo, T.~Vardanega, L.~Kosmidis, J.~Abella, E.~Mezzetti, E.~Q. nones, and F.~J. Cazorla, ``Measurement-based probabilistic timing analysis for multi-path programs,'' in \emph{Proceedings of the 24th Euromicro Conference on Real-Time Systems (ECRTS)}, 2012, pp. 91--101.

\bibitem{davis2019probabilistic}
R.~I. Davis and L.~Cucu-Grosjean, ``A survey of probabilistic timing analysis techniques for real-time systems,'' \emph{Leibniz Transactions on Embedded Systems}, vol.~6, no.~1, pp. 03:1--03:60, 2019.

\bibitem{burns2017survey}
A.~Burns and R.~I. Davis, ``A survey of research into mixed criticality systems,'' \emph{ACM Computing Surveys}, vol.~50, no.~6, pp. 1--37, 2017.

\bibitem{lelli2016deadline}
J.~Lelli, C.~Scordino, L.~Abeni, and D.~Faggioli, ``Deadline scheduling in the {Linux} kernel,'' \emph{Software: Practice and Experience}, vol.~46, no.~6, pp. 821--839, 2016.

\bibitem{qfabric2026}
\BIBentryALTinterwordspacing
A.~Bhattacharya and A.~Chakraborty, ``{QFabric}: Guarded cross-domain task recovery on a {Linux}--{RTOS} board,'' Zenodo, software release v1.1.0, 2026, source repository: \url{https://github.com/elixpo/qfabric}; Apache-2.0. [Online]. Available: \url{https://doi.org/10.5281/zenodo.23692354}
\BIBentrySTDinterwordspacing

\end{thebibliography}
